\documentclass{cidr-2020}

\usepackage{ifthen}

\usepackage{pgfplots}
 \usepackage{subcaption} 
\usepackage{graphicx}
\usepackage{enumitem}
\usepackage{wrapfig}
\usepackage{listings}
\usetikzlibrary{patterns}
\graphicspath{{figures/}}
\usepackage[utf8]{inputenc}
\usepackage{balance}
\usepackage{lastpage}

\newboolean{commentSwitch}
\setboolean{commentSwitch}{true}

\ifthenelse{\boolean{commentSwitch}}{
	\newcommand{\jonas}[1]{\textcolor{red}{JT: #1}}
	\newcommand{\sebastian}[1]{{\bf\color{orange} SB: #1}}
	\newcommand{\steffen}[1]{\textcolor{blue}{SZ: #1}}
	\newcommand{\volker}[1]{{\bf\color{red} Volker: #1}} 
	\newcommand{\philipp}[1]{\textcolor{green}{PG: #1}}     
	\newcommand{\ventura}[1]{\textcolor{blue}{VDM: #1}} 
	\newcommand{\HG}[1]{{\bf\textcolor{cyan}{HG: #1}}}
	\newcommand{\dimitrios}[1]{{\bf\textcolor{purple}{DG: #1}}}
	\newcommand{\ankit}[1]{{\bf\textcolor{yellow}{AC: #1}}}
	\newcommand{\eleni}[1]{{\bf\textcolor{purple}{ETZ: #1}}}
}{
	\newcommand{\jonas}[1]{}
	\newcommand{\sebastian}[1]{}
	\newcommand{\steffen}[1]{}
	\newcommand{\volker}[1]{}  
	\newcommand{\philipp}[1]{}
	\newcommand{\ventura}[1]{}
	\newcommand{\HG}[1]{}
	\newcommand{\dimitrios}[1]{}
	\newcommand{\ankit}[1]{}
	\newcommand{\eleni}[1]{}			
}

\let\subparagraph\relax 

\definecolor{c-scotty-lazy}{RGB}{217,95,2}
\definecolor{c-scotty-eager}{RGB}{27,158,119}
\definecolor{c-scotty-ff}{RGB}{117,112,179}
\definecolor{c-buckets}{RGB}{231,41,138}
\definecolor{c-flatfat}{RGB}{102,166,30}
\definecolor{c-flatfatp}{RGB}{230,171,2}
\definecolor{c-naive}{RGB}{166,118,29}
\definecolor{c-pairs}{RGB}{102,102,102}
\definecolor{c-cutty-lazy}{RGB}{102,166,30}
\definecolor{c-cutty-eager}{RGB}{230,171,2}
\definecolor{c-cutty-ff}{RGB}{150,150,150}
\definecolor{c-cutty-ff}{RGB}{150,150,150}
\definecolor{c-purple}{RGB}{161,13,89}

\newcommand{\plotfileNOMARK}[1]{
    \pgfplotstableread[col sep=tab]{#1}{\table}
    \pgfplotstablegetcolsof{\table}
    \pgfmathtruncatemacro\numberofcols{\pgfplotsretval-1}
    \pgfplotsinvokeforeach{1,...,\numberofcols}{
        \pgfplotstablegetcolumnnamebyindex{##1}\of{\table}\to{\colname}
        \addplot+[thick, mark=none] table [y index=##1] {#1}; 
        \addlegendentryexpanded{\colname}
    }
}

\definecolor{lst-gray}{rgb}{0.5,0.5,0.5}

\newcommand*\makecircled[1]{\tikz[baseline=(char.base)]{\node[shape=circle,draw,color=black,fill=white,inner sep=0.4mm,line width=0.2mm,text=black] (char) {\textbf{#1}};}}

\definecolor{Dark2-3-1}{RGB}{27,158,119}
\definecolor{Dark2-3-2}{RGB}{217,95,2}
\definecolor{Dark2-3-3}{RGB}{117,112,179}
\definecolor{Dark2-7-5}{RGB}{230,171,2}

\usetikzlibrary{colorbrewer}
\usepgfplotslibrary{colorbrewer}
\usepackage{url}

\begin{document}

  \newcommand{\fett}[1]{{\bf #1}}

\title{The NebulaStream Platform: Data and Application Management for the Internet of Things}

\author{
%
%
\alignauthor
Steffen Zeuch{$^{1,2}$}\hspace*{2mm}
Ankit Chaudhary{$^{1}$}\hspace*{2mm}
Bonaventura Del Monte{$^{1,2}$}\hspace*{2mm}
Haralampos~Gavriilidis{$^{1}$}\\
Dimitrios Giouroukis{$^{1}$}\hspace*{2mm}
Philipp M. Grulich{$^{1}$}\hspace*{2mm}
Sebastian Breß{$^{1}$}\hspace*{2mm}
Jonas Traub{$^{1,2}$}\hspace*{2mm}
Volker Markl{$^{1,2}$}\\
\affaddr{\vspace{4mm}$^1$Technische Universit{\"a}t Berlin \hspace{1cm} $^2$DFKI GmbH \hspace{1cm}}
}

\maketitle
\begin{abstract}
The Internet of Things (IoT) presents a novel computing architecture for data management: a distributed, highly dynamic, and heterogeneous environment of massive scale.
Applications for the IoT introduce new challenges for integrating the concepts of fog and cloud computing as well as sensor networks in one unified environment.
In this paper, we highlight these major challenges and outline how existing systems handle them.
To address these challenges, we introduce the NebulaStream platform, a general purpose, end-to-end data management system for the IoT.
NebulaStream addresses the heterogeneity and distribution of compute and data, supports diverse data and programming models going beyond relational algebra, deals with potentially unreliable communication, and enables constant evolution under continuous operation.
In our evaluation, we demonstrate the effectiveness of our approach by providing early results on partial aspects.

\end{abstract}
\section{Introduction}
\label{sec:intro}
Over the last decade, the amount of produced data has reached unseen magnitudes.
Recently, the International Data Corporation~\cite{dataAge} estimated that by 2025 the global amount of data will reach 175ZB and that 30\% of these data will be gathered in real-time.
In particular, the number of IoT devices is expected to grow to as many as 20 billion connected devices by 2025~\cite{gartner}.
At the same time, devices such as embedded computers or mobile phones continuously increase their processing capabilities.
This trend enables the exploitation of their computing and communication capabilities, as they become objects of common use.
As a result, the IoT is one of the fastest emerging trends in the area of information and communication technology~\cite{IOTVison}.

\begin{figure}[t]
    \includegraphics[scale=.2]{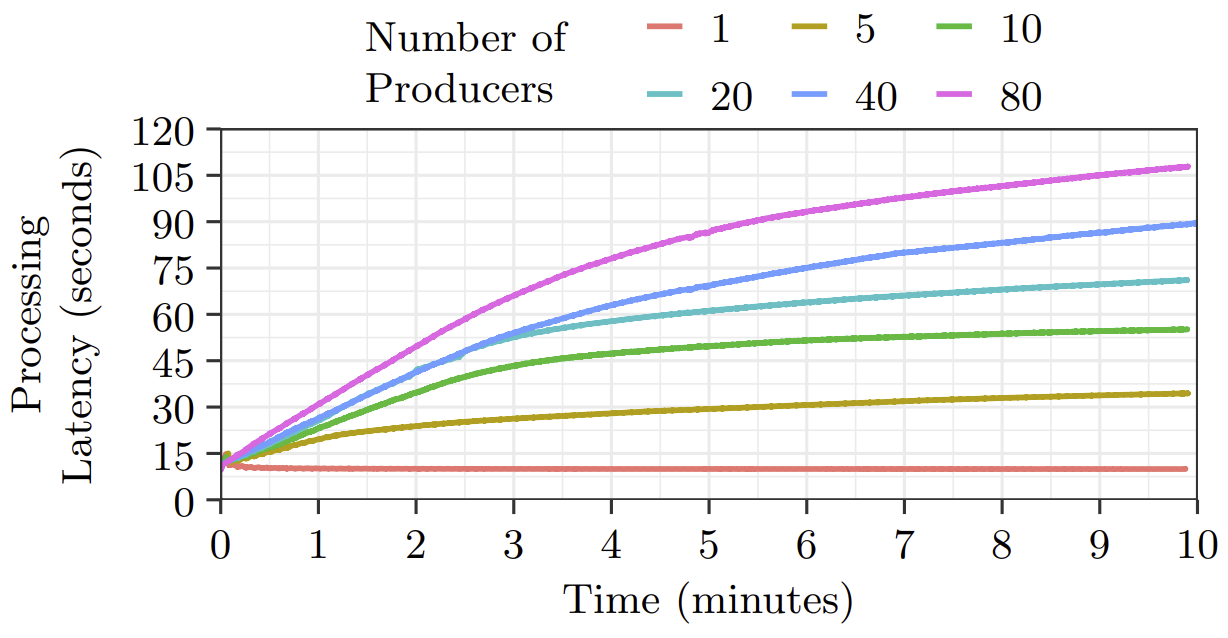}
    \vspace{-5mm}
	\caption{IoT application using a cloud-centric SPE.}
    \label{fig:throughput-latency}
    \vspace{-7mm}
\end{figure}

The explosion in the number of connected devices triggers the emergence of novel data-driven applications.
These applications require low-latency, location awareness, wide-spread geographical distribution, and real-time data processing on potentially millions of distributed data sources.
To enable these applications, a data management system needs to leverage the capabilities of IoT devices.

However, today's data management systems are not yet ready for these applications as they embrace either the cloud or the fog computing paradigm.
Systems based on the cloud paradigm, e.g., Flink~\cite{Stratosphere}, Spark~\cite{zaharia2016apache}, and Kafka Streams~\cite{kstreams2018}, do not exploit the full capabilities of IoT devices.
To implement IoT applications, these systems require the collection of sensor data centrally in a data center prior to applying processing.
This centralized processing paradigm presents a bottleneck for upcoming IoT applications, which need to process data from millions of distributed sensors.

In Figure~\ref{fig:throughput-latency}, we showcase the impact of this bottleneck by executing an IoT application scenario using a cloud-based approach and reporting the average processing latency.
To this end, we scale the number of IoT data producers from 1 to 80.
Each producer generates data at a constant speed of 50K record/sec.
Producers send their data over a gateway to an Kafka cluster with five nodes.
Inside the same cloud environment, we setup an Flink cluster with eight nodes (cloud nodes are connected through a 1 Gbit Ethernet connection).
Our Flink job reads data from Kafka and executes a tumbling windowed aggregation of 10 seconds to count distinct events.
We let the experiment run for 10 minutes and measure the end-to-end processing latency following the methodology introduced by Karimov et al.~\cite{KarimovRKSHM18}.
Our experiment shows that latency increases as we increase the number of producers. 
Our cloud-based IoT application scenario can sustain up to 20 producers with constant latency.
Beyond this point, our application saturates and latency increases gradually.
This effect intensifies for more IoT producers and results in a continuously increasing backlog within Kafka.
Overall, our experiment shows that a centralized cloud approach does not scale for IoT applications and thus future IoT applications require a new system.

In contrast, systems based on the fog computing paradigm, e.g., Frontier~\cite{Frontier} and CSA~\cite{Streaming_IOT_Survey}, exploit the processing capabilities of edge devices, i.e., devices that are physically closer to the data sources. 
These devices apply data reduction techniques, e.g., pre-selection or pre-aggregation, to reduce data volume as early as possible in the processing pipeline, i.e., close to the sensor.
However, fog computing systems only scale within the fog and do not exploit the virtually unlimited resources of modern cloud infrastructures (e.g., Amazon Web Services or Microsoft Azure).
 
Data management systems for wireless sensor networks (WSNs), e.g., TinyDB~\cite{tinyDB}, exploit small battery-powered sensors to create a network of nodes to capture physical phenomena, such as earthquakes or volcanic eruptions.
These systems apply acquisitional query processing techniques to optimize the execution for battery lifetimes and deploy a small set of specialized queries to capture the physical phenomena.
However, WSN systems only scale within the sensor networks and do not exploit the resources of the attached cloud and fog environments. 
In particular, they do not consider offloading computation to external nodes and do not provide general-purpose query execution capabilities.

Overall, there is no general-purpose, end-to-end data management system for a unified sensor-fog-cloud environment with functionality similar to production-ready systems such as Flink or Spark.
To enable future IoT applications, a data management system for the IoT has to combine the cloud, the fog, and the sensors in a single unified platform to leverage their individual advantages and enable cross-paradigm optimizations (e.g., fusing, splitting, or operator reordering).
From a system point of view, this unified environment imposes three unique characteristics that are not supported by state-of-the-art data management systems.

\textbf{Heterogeneity:} A unified environment consists of a high\-ly heterogeneous hardware landscape.
The processing nodes range from low-end battery-powered sensors (e.g., Mica \linebreak Motes) over system-on-a-chip devices (e.g., Raspberry PIs) to high-end rack-scale servers. 
In particular, cloud infrastructures consist of homogeneous node setups, whereas the fog contains heterogeneous, low-end computing devices.
Furthermore, WSNs consist of highly specialized battery-\linebreak powered sensors.
To exploit the individual capacities of each node, an IoT data management system has to take their individual capabilities into account, especially their resource restrictions.
However, current data management systems abstract from the underlying hardware with virtual machines and managed runtimes.
These abstractions hinder the exploitation of specialized instructions and processing units and prevent important optimizations.

\textbf{Unreliability:} A unified environment has to handle different levels of runtime dynamics.
The fog introduces a highly dynamic runtime environment with unreliable nodes that might change their geo-spatial position, i.e., resulting in many transient errors or changes in latency/throughput.
WSNs exacerbate this highly dynamic runtime even further by turning-off sensors temporally to save energy and allowing reads only following a dedicated read schedule.
In contrast, a cloud infrastructure is a relatively stable environment where node failures are rare.
However, current approaches for load balancing, fault-tolerance, and correctness only concentrate on one particular environment. Thus, these approaches miss out important cross-paradigm optimization potential.

\textbf{Elasticity:} In a unified environment, data move from the sensors via intermediate nodes to the cloud, and finally to the consumer, e.g., a user device or another system.
The fog topology is commonly built as a tree-like network topology~\cite{bonomi2012fog, hong2013mobile} with several dataflow paths.
Data processing in the fog topology has to be network-aware because only nodes on the path from the sensors to the cloud can participate.
Furthermore, in a WSN, all sensors send their data to the next sensor in range until all data end up at the root of the network.
In contrast, in the cloud, every node has access to all data, e.g., via a distributed file system, e.g., HDFS.
However, current approaches allow optimizations, scaling, and load balancing only within nodes of the same environment and thus miss out important cross-paradigm optimization potential.

Overall, a unified environment introduces a previously unprecedented, unique combination of characteristics, i.e., hardware heterogeneity, unreliable nodes, and changing network topologies.
This new set of characteristics enables new cross-paradigm optimizations, which are crucial to support upcoming IoT applications over millions of sensors.

In this paper, we propose \textit{NebulaStream} (NES), a novel data processing platform that addresses the above-mentioned heterogeneity, unreliability, and scalability challenges and enables effective and efficient data management for the IoT. 
In particular, NES copes with these unique characteristics as follows.
First, NES copes with heterogeneity by maximizing \textit{sharing of results} and \textit{efficiency of computing} to significantly reduce the amount of data transferred and to exploit hardware capabilities efficiently.
Second, NES addresses unreliability by applying \textit{dynamic decisions} and \textit{incremental optimizations} during runtime to be as flexible as possible.
Third, NES enables elasticity by designing each node to react \textit{autonomously} to a wide range of situations during runtime.
With NES, we enable future IoT applications by unifying sensors, fog, and cloud in one general-purpose, end-to-end data management platform.
Our early experiments show that NES reduces the amount of data and sensor reads up to 90\%, increases node throughput and decreases energy consumption on low-end devices by up to two orders of magnitude, and processes queries with low latency even in the presence of many node failures.

The remainder of the paper is structured as follows.
We show a typical IoT application scenario in Section~\ref{sec:iot_example_application}.
In Section~\ref{sec:neb_overview}, we describe the NebulaStream platform, discuss its design principles, and provide initial performance results.
Finally, we survey related work in Section~\ref{sec:sota} and conclude in Section~\ref{sec:conclusion}.

\section{I\MakeLowercase{o}T Application Scenario} 
\label{sec:iot_example_application}
\begin{figure}[t!]
	\centering
	\includegraphics[width=\linewidth]{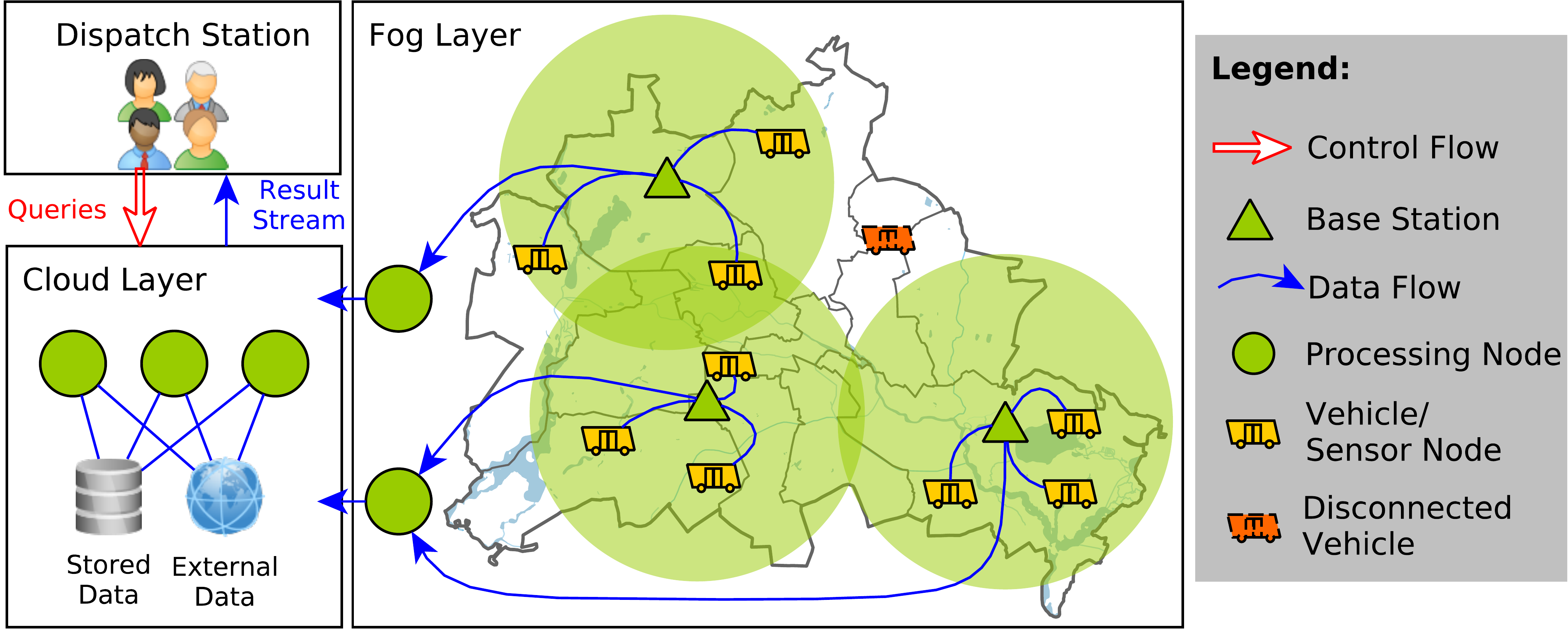}
		\vspace{-0.6cm}
	\caption{IoT application scenario.}
		\vspace{-0.6cm}
	\label{fig:iot-overview-berlin}
\end{figure}

In Figure~\ref{fig:iot-overview-berlin}, we present an integrated public transport system of Berlin as a representative IoT application scenario.
The components in this scenario are either stationary or mobile.
Vehicles (red and yellow boxes), i.e., taxis, buses, subways, and trains move around the city and carry a set of sensors and a simple processing unit.
Each unit collects vehicle data (e.g., routing, maintenance information, and occupancy/usage) as well as data from the environment (e.g., traffic, road conditions, and weather).
The base stations, processing nodes, and dispatch station are stationary components.
Base stations (green triangles) are distributed across the city and consist of antennas, network routers, and compute and storage capacity. 
Processing nodes (green circles) are distributed within the city to gather data from several base stations and apply more complex processing.
The centralized dispatch station represents the endpoint for all data and merges data from the fog and the cloud with stored and external data.
Users manage public transport through the dispatch station.
This IoT scenario requires a massively distributed system with continuous data producers as well as transient and permanent, distributed compute and storage capabilities.

The environment in this scenario differs fundamentally from current cloud-based data processing architectures.
In particular, vehicles move within the city and interact with multiple antennas, which transmit data to base stations.
Due to the dynamic nature, vehicles may encounter temporary connection losses or outages (red vehicle), e.g., when they are outside of transmission ranges.
Furthermore, all vehicles move at different speeds, on different roads/tracks, and are potentially equipped with different hardware.
User queries addressing only a subset of the vehicles do not require collecting all sensor data from all vehicles at every transmission interval.
This represents a major characteristic that is crucial for enabling large-scale IoT applications.
As a result, a fog requires continuous adaptation to a dynamic environment with respect to faults and changes in the availability, amount, type, capacity, and location of data and compute nodes.
Furthermore, on the sensor level, a system has to continuously adapt the sensor reads depending on a dynamic query workload.

Despite the distributed nature, it must be possible to manage the system through a centralized, global view and execute continuous as well as ad-hoc data analytics.
This includes the entire data analysis pipeline, from information extraction to integration and model building using machine learning, signal processing, and other advanced analytics.

From a user perspective, this system may assist the public transport dispatcher to schedule new vehicles or reroute vehicles in case of outages or increased passenger demand.
This results in a feedback loop that may change the physical fog architecture.
Furthermore, this architecture allows for enriching real-time data with external sources, e.g., air pollution measurements, event calendars, area crowdedness, or knowledge bases.
The characteristics of this application are representative for many IoT scenarios including Industry 4.0, smart homes, smart grids, smart cities, or participatory sensing applications.
\section{NebulaStream Platform}
\label{sec:neb_overview}
In this section, we present the NebulaStream (NES) platform.
First, we describe the common topology of IoT application scenarios and highlight its novelty (Section~\ref{sub:nes_topology}).
After that, we identify key design principles for an IoT data management system (Section~\ref{sub:design_principles}) and later describe how NES implements them (Section~\ref{sub:arch_overview}).
Finally, we discuss challenges for an IoT data management system and how NES addresses them (Section~\ref{subsec:nes_challenges}).
\begin{figure}[t]
	\centering
	\includegraphics[width=\linewidth]{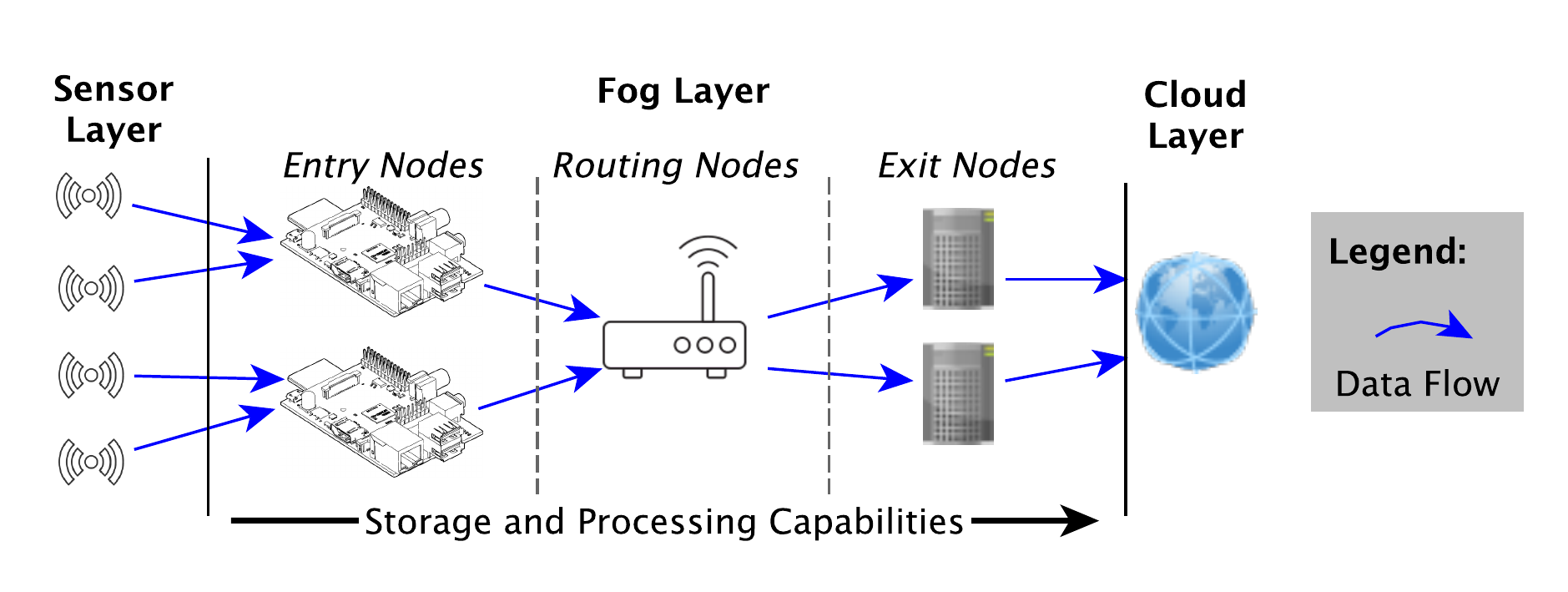}
	\vspace{-0.6cm}
	\caption{Multi-layer NES Topology.}
	\vspace{-0.6cm}
	\label{fig:overview}
\end{figure}
\subsection{NES Topology} 
\label{sub:nes_topology}
In Figure~\ref{fig:overview}, we present a multi-layer NES Topology that is common in today's IoT infrastructures~\cite{bonomi2012fog}.
This figure presents the dataflow from the sensors to the cloud.
The basic assumptions in this topology are three-fold.
First, all data might reach the \textit{Cloud Layer}.
Second, devices on the path from the sensors to the cloud are able to apply processing.
Third, the Cloud Layer is able to apply remaining processing, i.e., representing a fall-back mechanism.
In contrast, all other nodes can only access data if they are routed through them and their storage and processing capabilities determine the operations they can apply.

The data are routed among the three layers as follows.
On the \textit{Sensor Layer}, millions of sensors produce data without processing them.
However, NES is able to schedule the sensor reads depending on the query, e.g., increasing read frequency or omitting reads.
Sensors provide two data access patterns: pull-based and push-based.
Each sensor is connected to at least one low-end node in the \textit{Fog Layer}, which is responsible for this sensor (so-called \textit{Entry Node}).
In the \textit{Fog Layer}, NES processes data as they flow from Entry Nodes to Exit Nodes. During processing, nodes may change their geo-spatial position.
The data transfer is orchestrated by \textit{Routing Nodes}, such as routers or switches.
The data processing capabilities on Routing Nodes are restricted and the provided functionality is highly vendor-dependent~\cite{DPI,lerner}.
In general, the storage and processing capabilities of nodes increase significantly in the NES Topology with each hop towards the Cloud Layer.
After leaving the Fog Layer through an \textit{Exit Node}, data enter the Cloud Layer.
The Cloud Layer provides virtually unlimited scaling of compute and storage.
In IoT application scenarios, this layer will perform the remaining computation and output the data to the user.
An alternative approach to this centralized design would allow each node in the fog to function as a potential sink.
Thus, users would submit their queries directly through their device and each device would represent an exit node in the topology.
In this decentralized design, each device will be responsible for answering the submitted user query.
This design naturally supports geo-spatial query processing as most users are potentially only interested in data produced nearby.
Exploring the design space of a centralized vs. a decentralized design is one major future challenge. 

The NES Topology introduced in Figure~\ref{fig:overview} represents a fundamentally new and unique set of characteristics and requirements compared to common cloud infrastructures.
First, query processing and operator placement have to be network-aware.
The main query optimization goal is to find an efficient route through the Fog Layer that reduces data volumes as early as possible without violating any Service-Level-Agreement (SLA) but fulfilling Quality of Ser-\linebreak vice (QoS) constraints.
Second, the NES Topology is highly heterogeneous and many nodes have only limited processing capabilities.
In particular, nodes in the lower parts of the Fog Layer are restricted in storage and processing capabilities.
Furthermore, processing has to trade-off between energy consumption and performance.
Third, the Fog Layer is highly unreliable compared to the homogeneous and relatively stable Cloud Layer.
To support mobility and related aspects, the system has to take the characteristics of each individual environment into account.
Fourth, the volume and velocity of sensor data represent an external factor.
As a result, the entire system has to evolve around sensor data that is injected by the outside world.
With \mbox{NES}, we build a platform that creates a federation of sensors, fog, and cloud, which enables big data acquisition and analysis.
\begin{figure}[t]
	\centering
	\includegraphics[width=0.95\linewidth]{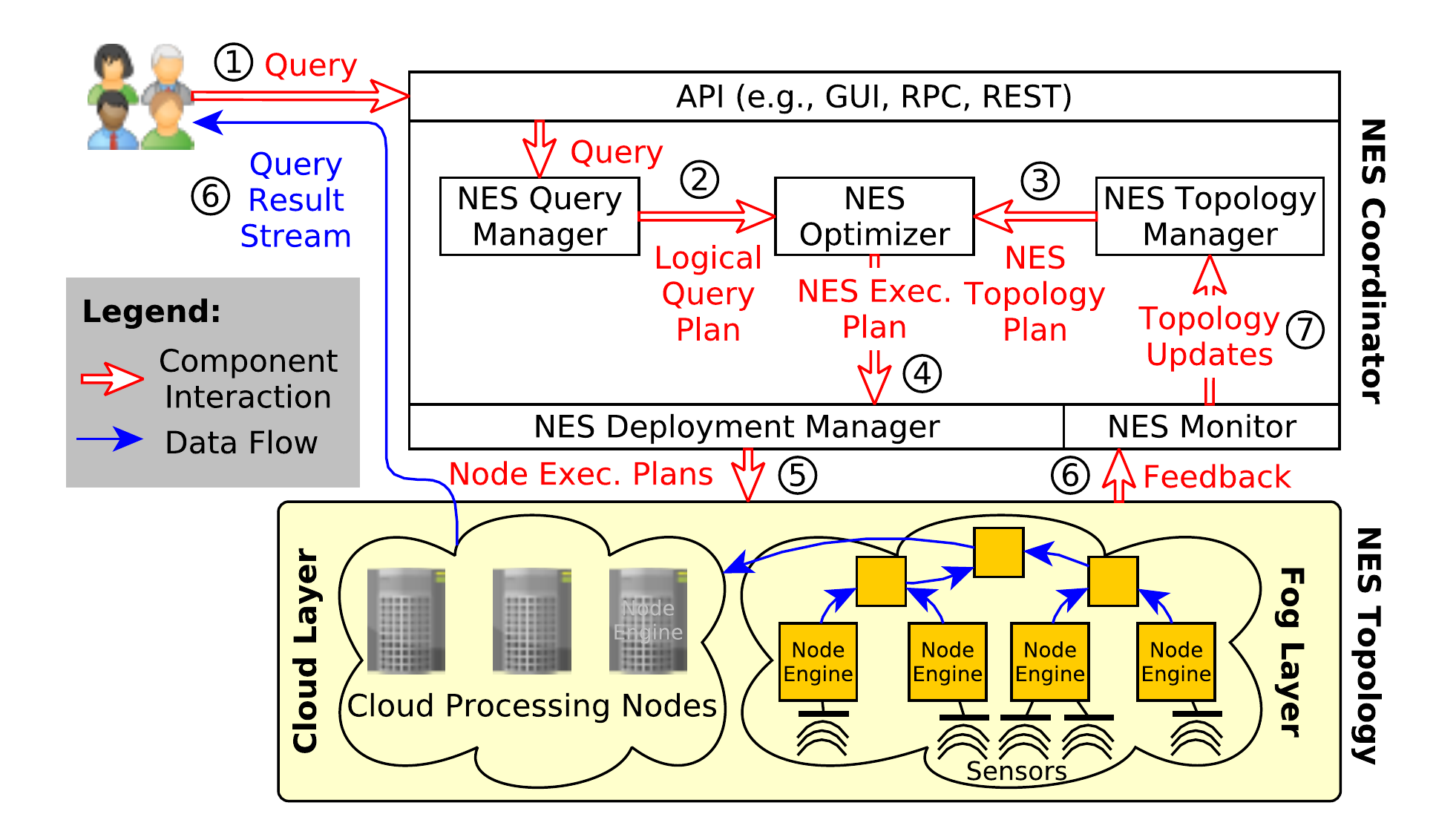}
	\vspace{-0.4cm}
	\caption{NES architecture overview.}
	\vspace{-0.5cm}
	\label{fig:sw_overview}
\end{figure}
\subsection{NES Design Principles} 
\label{sub:design_principles}
NES is a platform for future IoT applications that copes with the unique set of characteristics of a unified environment.
For individual layers, different approaches were proposed over the last decades.
However, combining all of them into a single system is the major challenge that we address with NES.
To handle millions of sensors and thousands of queries, we base the system design of NES on the following design principles:
\begin{enumerate}[topsep=4pt, itemsep=-4pt, leftmargin=*] 
	\item \textbf{Dynamic Decisions:} NebulaStream never expects a static behavior or conditions in any component.
	\item \textbf{Autonomous Processing:} NebulaStream equips compute nodes with all logic necessary to act as autonomously as possible.
	\item \textbf{Incremental Optimizations:} NebulaStream optimizes a network of active queries in incremental steps rather than traditional query optimization or batched changes.
	\item \textbf{Maximize Sharing:} NebulaStream shares data and processing wherever possible, i.e., on windows (stream slicing), among queries (multi-query optimization), on sensor data (acquisitional query processing), and on operator level (code optimization).
	\item \textbf{Maximize Efficiency:} NebulaStream applies hardware-tailored code generation to exploit the underlying hardware efficiently.
	\item \textbf{SLA Centric Processing:} NebulaStream's primary goal is to match user-provided SLAs and QoS constraints with available resources.
	\item \textbf{Ease of Use:} NebulaStream enables users to choose their preferred programming environments and models, without worrying about system-internals and performance implications. 
\end{enumerate}

\subsection{NES Architecture} 
\label{sub:arch_overview}
In Figure~\ref{fig:sw_overview}, we present the architecture of \mbox{NebulaStream}.
In general, we design NES with a centralized deployment process and a decentralized run-time re-optimization. 
In particular, we envision a \textit{logically} centralized deployment process in which one central instance has control over the deployment.
However, this logically centralized instance can be distributed among multiple region coordinators to form a hierarchy of coordinators.
In the future, we envision moving towards a decentralized deployment process that enables every device to timely submit queries and receive results.
In the current design, users interact with NES through one of the provided APIs to send queries to the \textit{NES Coordinator}~\makecircled{1}.
Our current APIs allow specifying dataflow programs, similarly to the APIs of streaming systems like Flink, Spark, and Storm. 
The NES Coordinator consists of several components that orchestrate query processing.
The \textit{NES Query Manager} is responsible for creating logical query plans from user requests~\makecircled{2}. 
Additionally, this component maintains \textit{logical streams} that represent logical views over sensors, e.g., a logical stream \textit{cars} could combine sensor inputs from multiple cars into one consistent stream.
The \textit{NES Topology Manager} orchestrates the NES Topology, which consists of workers and sensors.
During startup, each device registers itself and provides information, such as resource capabilities and network topology information.
However, to reduce the complexity of optimization decisions, NES follows the idea of introducing \textit{zones} that aggregate a sub-tree or geo-spatial region of the topology into one node.
Thus, the optimizer treats a zone as one node which transparently abstracts from the dynamic behavior inside the zone.
As a result, a topology may consist of a hierarchy of zones, which simplifies the global optimization process.
The efficient assembly of zones is one future research challenge for NES.
 
The \textit{NES Optimizer} provides the assignment of a logical query plan (created by the NES Query Manager) to the current NES Topology plan~\makecircled{3} (maintained by the NES Topology Manager).
This assignment defines the \textit{NES Execution Plan (NES-EP)}.
The assignment process introduces a large optimization search space, e.g., operators can be assigned top-down, bottom-up, or by other assignment strategies.
The \textit{NES Deployment Manager} takes the NES-EP~\makecircled{4}, disassembles it into Node Execution Plans (Node-EPs), deploys them to the nodes in the NES Topology, i.e., into either the Fog or the Cloud Layer, and sets up the sensors~\makecircled{5}.
This deployment is performed incrementally and requires rerouting data on different dataflow paths.
Note that this deployment process has to handle a gap between optimization and deployment time. 
Thus, optimization is based on a snapshot of the topology, while deployment has to take the current topology into account.
Therefore, the deployment process in this highly dynamic execution environment introduces many interesting research challenges, such as the partial deployment of plans and the partial re-optimization of sub-plans.
The \textit{NES Monitor} constantly collects feedback from the NES Topology~\makecircled{6} and maintains statistics and current resource utilization for the NES Topology Manager~\makecircled{7}.
To improve operator placement, the NES Optimizer requests these statistics and current resource utilization from the \textit{NES Monitor}~\makecircled{7}.
However, maintaining a centralized, coherent view over a large and highly dynamic topology is a major research challenge.
First, the NES Optimizer has to be aware that the topology data is potentially out-dated and thus has to optimize accordingly, e.g., by providing a set of alternative plans.
Second, the collection of monitoring data and the maintenance of statistics has to take the current system load into account and thus must be prioritized lower than data transfers to answer user queries.
Third, we envision a decentralized run-time re-optimization process that is triggered by the nodes themselves.
To this end, NES nodes first attempt to address a change locally, then communicate with their neighboring nodes, and finally requesting support from a central coordinator.

\begin{figure}[t]
	\centering
	\includegraphics[width=\linewidth]{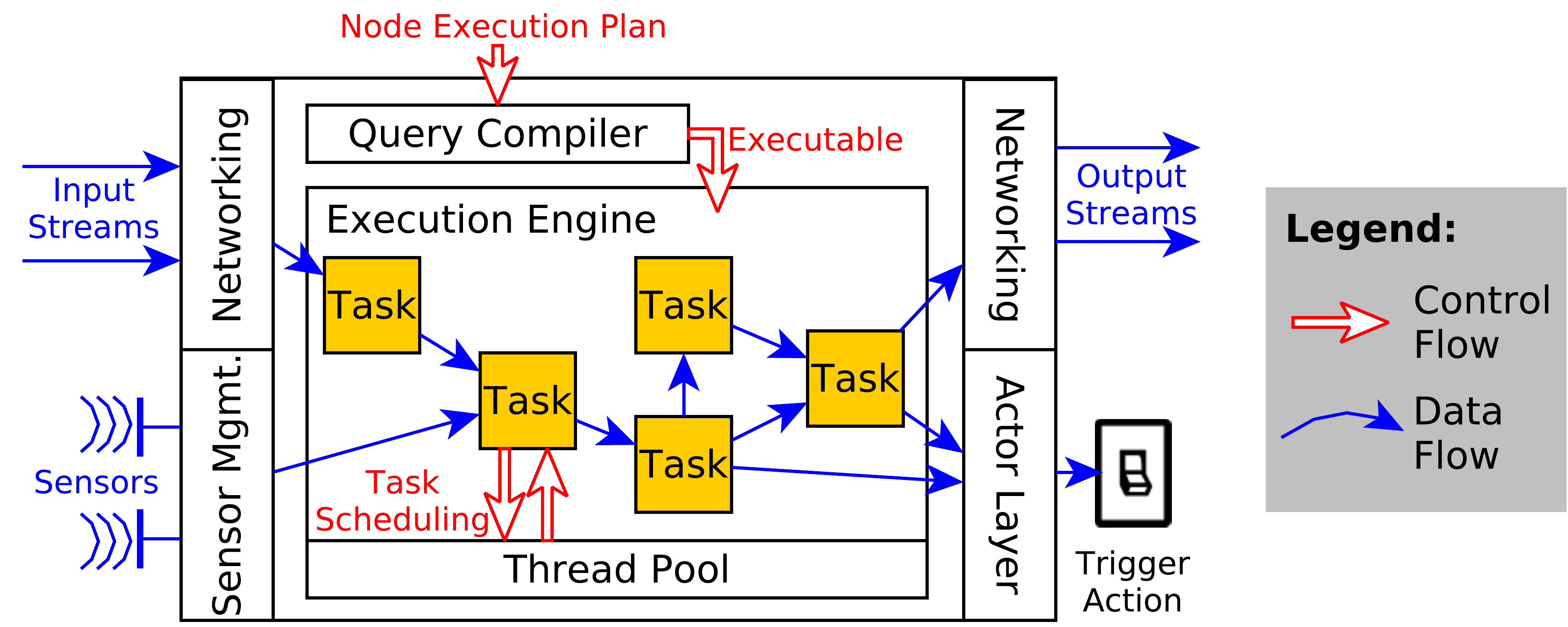}
	\vspace{-0.5cm}

	\caption{NES Node Engine.}
		\vspace{-0.5cm}

	\label{fig:node_overview}
\end{figure}

In Figure~\ref{fig:node_overview}, we show the components of the node engine, which is deployed on all devices of the NES Topology.
The \textit{NES Node Engine} is responsible for communicating with the NES Coordinator, accepting Node-EPs and control messages, as well as setting up the input sources, output sinks, and other components.
The incoming queries are Node-EPs, which contain a partial subtree of the overall NES-EP. 
The Node-EP is compiled by the local query compiler and later injected into the processing tasks.
As input, the NES Node Engine receives data from the network, e.g., from another node, or directly from an attached sensor.
As output, the NES Node Engine either sends data over the network or triggers an action on an attached device, e.g., controlling an actuator such as a light switch.

The \textit{Execution Engine} orchestrates the processing inside each NES  Node Engine.
The central unit of work is one task that combines $n$ input buffers, $m$ output buffers, and the execution of the specified operators~\cite{QTM}.
The processing in NES is \textit{source-driven} and applies the following sequence of steps on each incoming buffer.
First, the engine assembles the tasks by embedding the executable and allocates all required input, intermediate, and output buffers.
After that, the engine enqueues the tasks in one of the processing queues.
Finally, each thread in the \textit{Thread Pool} dequeues one task, processes it, and either enqueues the result buffer into an output queue or triggers an action.
This highly dynamic design enables high resource utilization but also introduces a dynamic execution order, which poses new challenges for the system design.

In addition to processing components, each NES Node Engine contains dedicated components for local and neighboring optimizations, windows, routing, sensors, state, and run-time re-optimization.
As a result, we drastically reduce the complexity of the query compiler and increase maintainability and separation of concerns in NES.
In particular, NES compiles only the \textit{hot} code fragments and links other functionalities as pre-compiled components (following Neumann et al.~\cite{neumann2011efficiently}).
Overall, it is a design decision in NES to equip the NES Node Engine with all necessary components to enable it to be as autonomous as possible.
In particular, we assign all means to the node to enable it to make as many decisions as possible decentrally and independently.
This design follows the Borealis design~\cite{borealis} and tries to encounter transient changes locally and permanent changes globally.
In NES, we envision a system design with autonomous nodes and a simple coordinator to mitigate potential bottlenecks in large scale environments.

\subsection{NES Solutions for IoT Challenges}
\label{subsec:nes_challenges}
Based on the unique characteristics highlighted in Section~\ref{sec:intro} and IoT application scenarios presented in Section~\ref{sec:iot_example_application}, we outline five main challenges for an IoT data management system. In the following, we discuss the challenges and propose our solutions.

\subsubsection{C1 - Heterogeneity, Distribution, and Volume of Data At-Rest and Data In-Motion}
NebulaStream's goal is to scale to thousands of queries and millions of sensors.
In the IoT, data are generated by many distributed sources such as sensors or streams of other systems.
A particular challenge originates from handling the sheer amount of diverse data sources, potentially up to the number of millions.
These sources differ in their characteristics, ranging from millions of small sensor streams to a few large streams from sources such as click-streams or auctions.
The accessibility of sources under security and privacy constraints, as well as efficient access paths, requires solutions completely different from what today's big data processing systems provide.
For example, an IoT infrastructure enables new solutions for security and privacy as it allows local pre-processing of data next to the generation, e.g., inside a house or building.
This enables a scenario where only authorized or anonymized data are sent to the central cloud.
As a result, we can enable users to have full control of their own data.
Overall, these characteristics imply research questions with respect to scalability, efficiency, integration, security, privacy, and interoperability.

To support this extreme diversity in NES, we follow the \textit{Maximize Sharing} design principle (Section~\ref{sub:design_principles}) and apply data sharing techniques on three different levels.
First, on the query level, NES exploits data sharing among multiple streaming queries as proposed by Karimov et al.~\cite{AStream}.
Second, on the operator level, NES slices data streams and exploits data sharing on stream aggregations as proposed by Traub et al.~\cite{GeneralStreamSlicing}.
Third, on the sensor level, NES applies \textit{Acquisitional Query Processing} (ACQP)~\cite{tinyDB} and \textit{On-Demand Scheduling} of sensor reads and data transmissions~\cite{OnDemandDataAcc}. These techniques limit data acquisition to data points which are required for answering user queries.
By combining the introduced techniques in NES, we attempt to drastically reduce the amount of acquired, transferred, and processed data; thus, enabling IoT applications with thousands of queries over millions of sensors.

\begin{figure}[t]%
    \centering%
	\includegraphics[scale=.21]{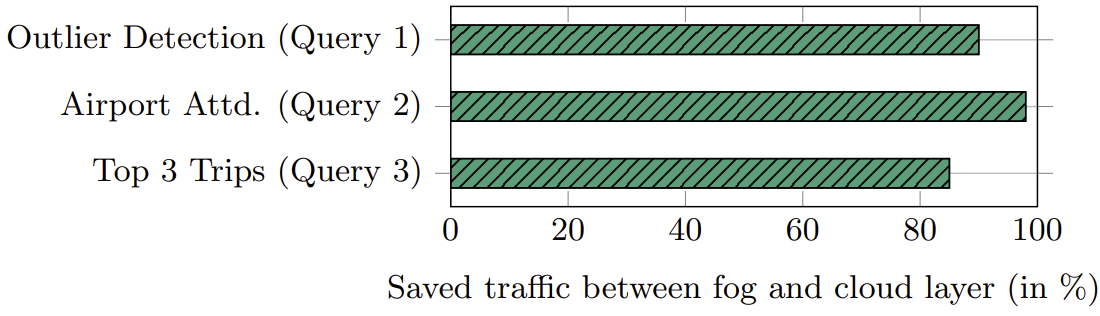}%
    		\vspace{-0.2cm}

    \caption{NES data reduction on the sensor level.}%
    		\vspace{-0.5cm}

    \label{fig:sense_example}%
\end{figure}

Figure~\ref{fig:sense_example} presents an initial experiment that demonstrates the potential savings of data reduction techniques in NES on the sensor level.
We use the New York taxis data set~\cite{nyt}, derive routes for each taxi trip, and replay the routes of all taxis on Raspberry Pis, which represent sensor nodes located in taxis.
As a baseline, we use a common IoT setup where sensor nodes stream current values to a central SPE in the cloud, without any knowledge about the executed queries.
In contrast to this cloud-centric IoT setup, NES combines cloud and fog nodes as well as sensor nodes in taxis in one system to allow for holistic optimizations.

We show three example queries in an SQL-like notation.
The queries include an outlier detection (Query~\ref{lst:q1}), an airport attendance monitoring (Query~\ref{lst:q2}), and a top three query for the longest ongoing trips  (Query~\ref{lst:q3}).\\

\definecolor{codegreen}{rgb}{0,0.6,0}
\definecolor{codegray}{rgb}{0.5,0.5,0.5}
\definecolor{codepurple}{HTML}{C42043}
\definecolor{backcolour}{HTML}{F2F2F2}
\definecolor{bookColor}{cmyk}{0,0,0,0.90}  
\color{bookColor}

\lstdefinestyle{mystyle}{
    backgroundcolor=\color{backcolour},   
    commentstyle=\color{codegreen},
    keywordstyle=\color{codepurple},
    numberstyle={},
    stringstyle=\color{codepurple},
    basicstyle=\footnotesize\ttfamily,
    breakatwhitespace=false,
    breaklines=true,
    captionpos=b,
    keepspaces=true,
    numbers=left,
    numbersep=2pt,
    showspaces=false,
    showstringspaces=false,
    showtabs=false,
}
\lstset{style=mystyle}

\lstset{emph={%
    AHEADLIMIT, DELAYLIMIT%
    },emphstyle={\color{codepurple}\bfseries}%
}%

\renewcommand{\lstlistingname}{Query}

\begin{lstlisting}[
	language=sql,
	numbers=none,
	caption={Journeys leaving the New York area and journeys without passengers. Checked every 2 seconds.},
	label=lst:q1,
	captionpos=b]
SELECT ts, medallion, trip_id, latitude, longitude, distance, passenger_count
FROM stream(taxis, 2000)
WHERE journey_flag=TRUE &&
    (latitude<40.249448 || latitude>41.381560 || longitude<-74.820611 || longitude>-71.848319 || distance=0 ||passenger_count=0); -- NY area
\end{lstlisting}

\begin{lstlisting}[
	language=sql,
	numbers=none,
	caption={Returning the number of passengers in the airport zone. Updated every 5 seconds.},
	label=lst:q2,
	captionpos=b]]
SELECT ts, sum(passenger_count)
FROM stream(taxis, 5000)
WHERE (40.536532<latitude AND latitude<40.745906)  && (-73.946390<longitude AND longitude<-73.609759) --airport
GROUP BY ts AHEADLIMIT 100 DELAYLIMIT 100; 
\end{lstlisting}

\begin{lstlisting}[
	language=sql,
	numbers=none,
	caption={Returning the top three longest ongoing trips. Updated every second.},
	label=lst:q3,
	captionpos=b]]
SELECT ts, latitude, longitude, trip_distance
FROM stream(taxis, 1000) 
WHERE journey_flag = TRUE
ORDER BY trip_distance DESC LIMIT 3;
\end{lstlisting}

\color{black}

We modify the data acquisition process for all three queries such that only required data are sampled and transmitted.
In particular, we can interleave data gathering operations (i.e., sensor reads) with data processing (e.g., filters)~\cite{tinyDB}.
Theoretically, the system has to read all sensors specified in the \textit{select} clause at the frequency specified in the \textit{from} clause.
However, the filter predicates in the \textit{where} clause allow for preventing sensor reads and data transmissions for tuples that are filtered out.
For instance, in Query~\ref{lst:q1} and Query~\ref{lst:q3}, we first check the journey flag.
If the value is \texttt{false}, we do not read any other sensor.

Another important optimization is to adjust sampling \linebreak rates continuously and to prevent data transmissions based on the observed sensor values~\cite{OnDemandDataAcc}.
For example, in Query~\ref{lst:q1} and Query~\ref{lst:q2}, we can use the current position of the taxi to calculate the earliest time when the taxi could leave New York or enter the airport area.
Thus, we know upfront that no tuple will pass the filter for that time span and do not have to read or evaluate sensor values for that time.
In addition, in Query~\ref{lst:q2}, we specify a tolerance for sensor read times (\textit{ahead} and \textit{delay} limit), which saves data transmissions when multiple queries request values from the same sensor.
We apply user-defined sampling functions to adjust sampling rates continuously, apply read time tolerances, and schedule sensor reads, respectively~\cite{OnDemandDataAcc}.
In Figure~\ref{fig:sense_example}, we show that the saved traffic between the fog and the cloud layer is significant for all queries using these optimizations.

\subsubsection{C2 - Heterogeneity, Distribution, and Volume of Compute}
NebulaStream's goal is to exploit the hardware resources of millions of heterogeneous devices efficiently.
A particular challenge originates from the potentially millions of compute devices that are found in a fog topology.
These devices have a diverse set of capabilities, with respect to storage, processing, and interconnect.
The devices range from small battery-powered sensors with no compute capabilities (beyond simple filtering) and an unreliable temporary connection  to a large compute cluster with huge storage, infiniband interconnect, and thousands of compute cores.
These characteristics imply challenges with respect to security, permission management, and efficient and effective resource utilization.

\begin{figure}[t]
	\centering
	\begin{subfigure}[c]{0.4\linewidth}
        \centering
		\includegraphics[scale=.21]{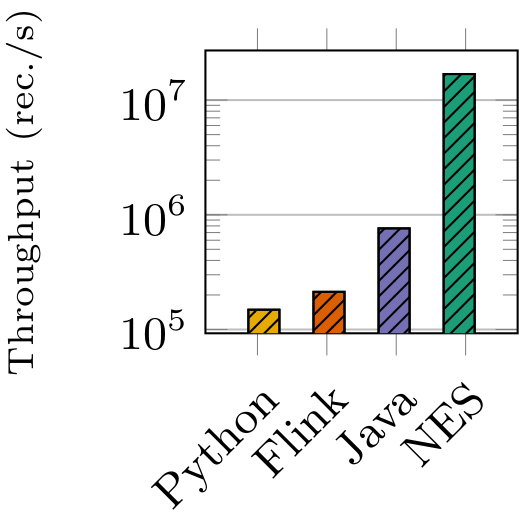}%
		\subcaption{Throughput.}
		\label{fig:pi_eval_throughput}%
	\end{subfigure}
	    \hfill
	\begin{subfigure}[c]{0.5\linewidth}
		\centering
		\includegraphics[scale=.18]{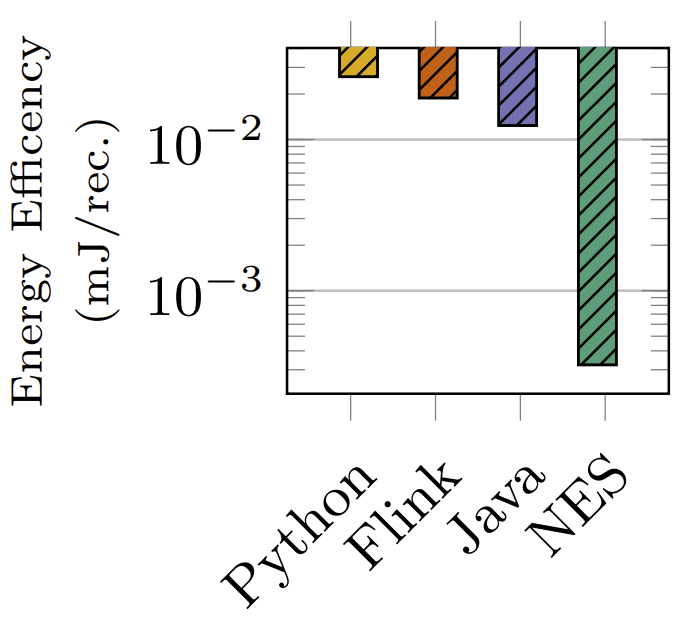}%
\vspace{-1mm}
		\subcaption{Energy Efficiency.}
		\label{fig:pi_eval_energy}%
\end{subfigure}
		\vspace{-0.3cm}
\caption{YSB on RaspberryPi 3B+.}%
		\vspace{-0.3cm}
\label{fig:pi_eval}%
\end{figure}

To support this heterogeneity in NES, we follow the \textit{Maximize Efficiency} design principle (Section~\ref{sub:design_principles}).
In particular, we apply two techniques.
First, we use query compilation, the leading paradigm for achieving high resource utilization in data-at-rest processing \cite{neumann2011efficiently}.
In NES, we transfer this approach to the special semantics of fog and stream processing.
In particular, NES generates specialized code depending on the actual query, hardware, and data characteristics~\cite{zeuch2019analyzing}.
Second, NES distributes query optimization and code generation between the central coordinator and the local node engine.
On the coordinator, NES performs global query optimizations (e.g., operator reorder) and splits the query into segments for individual devices.
On the node engine, the query compiler produces hardware-tailored code to exploit the availability capabilities most efficiently.

Our experiment in Figure~\ref{fig:pi_eval} evaluates the throughput and energy efficiency of the Yahoo Streaming Benchmark (YSB) on a RaspberryPi 3B+ using Python, Flink, a hand-opti\-mized Java program, and NES, respectively.
The YSB simulates a real-word stream processing task and consists of a filter and a windowed aggregation~\cite{yahooB}. 
We implement the YSB with a one second tumbling window and 10000 campaigns based on the codebase provided by Gier et al.~\cite{grier2016extending}. 
Our results show that hardware-tailored code generation is essential to efficiently utilize resources, especially for low-end devices.
In Figure~\ref{fig:pi_eval_throughput}, we present the maximal throughput of the four different YSB implementations.
NES outperforms all other systems by at least 10x and is the only system that is able to reach a throughput of more than 10 million tuples per second.
All other systems suffer from the high-overhead of the underlying managed runtime.
This overhead is significant on low-end devices like the RaspberryPi.
Furthermore, through code generation, NES reduces the energy consumption per device and thus requires less energy to achieve the same performance.
In Figure~\ref{fig:pi_eval_energy}, we evaluated the energy efficiency of the four different YSB implementations.
To this end, we define energy efficiency as the required energy in milli joule per processed record.
Our results show that NES requires around 0.0003 milli joule per tuple, which is an 80x improvement compared to the Python implementation.
In the future, we will further investigate the trade-off between energy  consumption and performance as one major research question for NES.
Especially for battery-powered sensors, code generation enables a higher operation time and thus reduced maintenance and replacement costs.

 \begin{figure}[t]
   \begin{center}%
     \includegraphics[scale=.2]{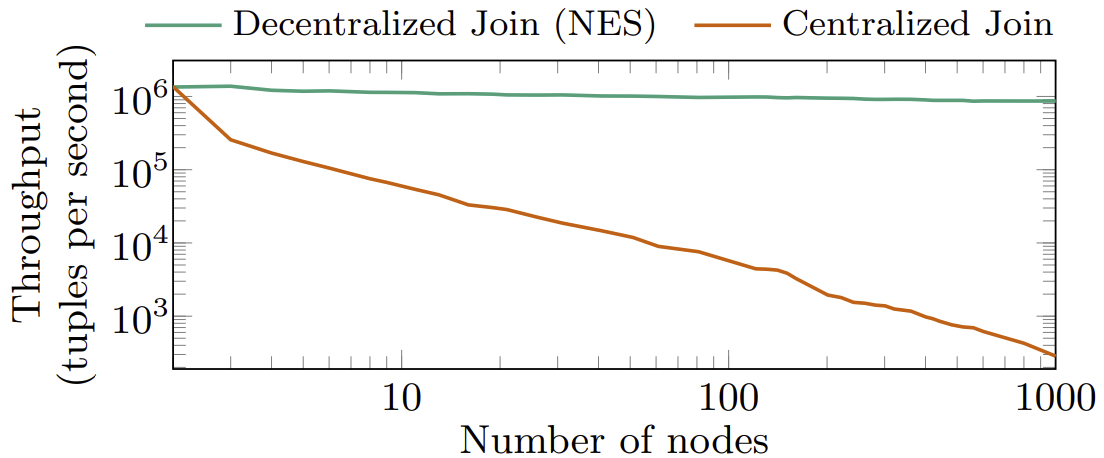}
     \vspace{-2.5mm}
     \caption{Gathering coherent snapshots from sensor nodes.}%
     \vspace{-9mm}
 	\label{fig:scalability_example}%
   \end{center}%
 \end{figure}

As a second technique, we utilize in-network processing inside the Fog Layer to reduce the computation required at the Cloud Layer.
In Figure~\ref{fig:scalability_example}, we present an example query that gathers values from up to 1000 nodes and joins them to coherent snapshots.
A snapshot is coherent, if all sensor values contained in the snapshot have been read at the same time.
In practice, snapshots are often incoherent, because the times of sensor reads are not perfectly aligned among all distributed nodes.
In addition, clock deviations among sensor nodes lead to undetected incoherence, which potentially causes application failures such as false correlations. 

We use the techniques which were introduced in the \linebreak SENSE System~\cite{traub2019sense} to ensure scalability and to mitigate incoherence. 
SENSE arranges sensor nodes in data gathering pipe\-lines, which join tuples incrementally (decentralized join) and ensure coherence.
In contrast to a centralized join, the Cloud Layer only joins the results of the pipelines instead of all individual sensor measures.
This prevents a central bottleneck at the Cloud Layer and ensures high throughput when gathering values from a large number of sensors.
As shown in Figure~\ref{fig:scalability_example}, a centralized join causes a drastic throughput decay when the number of nodes increases.
In contrast, by utilizing the available computing resources on the path from the sensors to the Cloud Layer, NES achieves almost constant throughput and addresses coherence issues.

By applying hardware-tailored code generation and in-network processing, NES exploits the available compute resources most efficiently and allows for balancing computational demands and energy consumption.

\subsubsection{C3 - Spontaneous, Potentially Unreliable \\ Connectivity between Data and Compute}
NebulaStream's goal is to detect and compensate potentially unreliable nodes in the Fog and Sensor Layer without impacting consistency and availability.
A particular challenge originates from the need to manage data and compute together, as most applications will consist of ad-hoc or standing streaming queries.
Furthermore, some compute units may be connected via Wifi, mobile, or satellite networks with intermittent connectivity and unreliable connections.
In contrast to a homogeneous and relatively stable cloud environment, a heterogeneous and volatile fog environment has to handle frequent transient failures.
Furthermore, WSNs are even more prone to transient failures due to their battery-powered low-end devices and vulnerable radio transmission.

Failures in the fog and in WSNs occur due to numerous reasons, most notably hardware errors, software errors, congestion that results in back-pressure (straggler nodes), inadequate resource allocation, and transient connection loss.
Furthermore, devices continuously refresh their connections while moving and create ad-hoc connections that result in an unpredictable communication pattern \cite{IOTVison}.
This requires special solutions to deal with the intermittent availability of resources, both with respect to data and code management.
The resulting challenges require changes in areas such as adaptivity, synchronization across devices, consistency, transaction management, recovery, and fault-tolerance.

 \begin{figure}[t]%
 	\centering
	 \includegraphics[scale=.2]{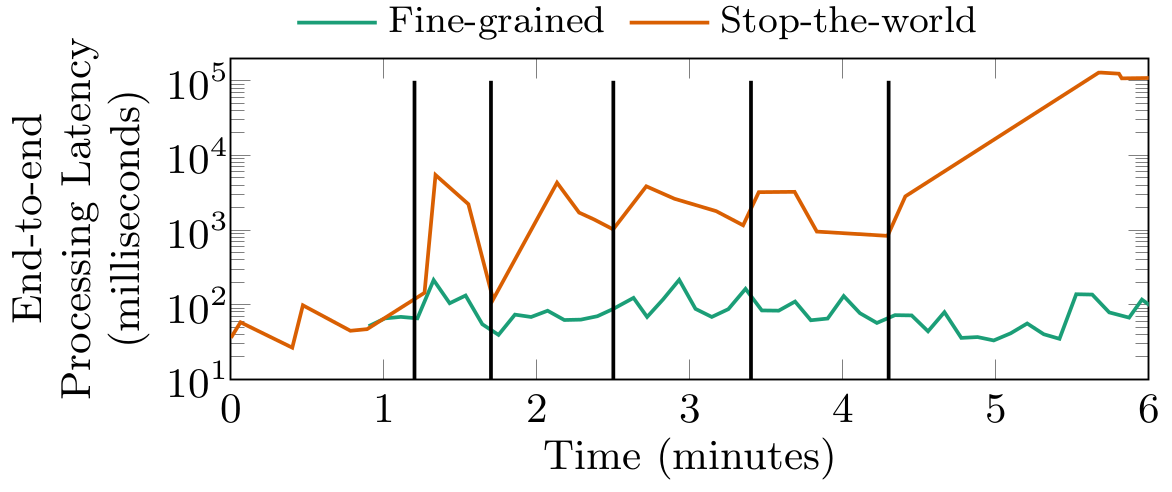}
     \vspace{-3mm}
     \caption{Evaluation of fault tolerance mechanisms.}
     \vspace{-7.5mm}
     \label{fig:failure_example}%
 \end{figure}

Common cloud-centric SPEs handle node failures using a stop-the-world recovery protocol~\cite{carbone2017flink,DBLP:reference/db/BalazinskaHS18}.
When an error occurs, the system stops the entire processing and redeploys a new query plan.
In contrast, NES adopts a fine-grained recovery protocol, i.e., NES restarts only the operator instances involved in a failure.
To assess the performance of both protocols, we implement them in Flink and run the comparison on a simulated IoT environment.
This environment comprises of 8 servers, which are equipped with Intel Xeon E5620 CPUs, 32 GB of RAM, and an 1 Gbits network.
In Figure~\ref{fig:failure_example}, we show the end-to-end processing latency of both protocols while randomly terminating compute nodes (indicated by the black vertical lines).
As shown, the stop-the-world protocol cannot recover from high transient error rates as the latency constantly increases.
In contrast, the fine-grained recovery protocol restarts failed operators without halting the entire query.

To achieve reliability in an unreliable environment, we apply the \textit{Dynamic Decisions}, \textit{Autonomous Processing} and \textit{Incremental Optimizations} design principles in NES.
Because a central component cannot keep up with the pace of failures in a dynamic environment, we apply a diverse set of techniques in NES.
On each layer of the NES Topology, we apply different failure recovery approaches; thus, providing different guarantees.
On the Sensor Layer, NES substitutes missing sensor values from broken sensors with nearby sensors, if applicable, or buffers the values during transient connection loss \cite{tinyDB}.
On the Fog Layer, NES extends the Frontier approach~\cite{Frontier}, which sends data through multiple network paths to achieve fault-tolerance.
Furthermore, data are buffered by upstream operators and replayed in case of an error.
On the Cloud Layer, NES extends existing fault-tolerance approaches, e.g., global checkpointing and message broker with fine-grained operator reconfiguration~\cite{VenturaPhDWork}.

By extending and combining existing approaches on different levels of the NES Topology into a unified fault-tolerant solution, we attempt to handle spontaneous, potentially unreliable connectivity of IoT infrastructures.

\subsubsection{C4 - Diversity in Programming and \\ Management Environments}
NebulaStream's goal is to support a diverse set of data processing workloads specified in different query languages and following different processing models (e.g., relational, linear, or graph algebra).
A particular challenge originates from IoT applications that require a combination of different data-oriented programming paradigms.
Possible workloads range over the entire data management pipeline, from information extraction over information integration to model building and inference.
In particular, running AI/ML/Data Science algorithms in the fog enables direct feedback loops between the digital and the physical world.
These workloads include potentially iterative algorithms mixing relational, linear, and graph algebra, and may run on top of continuous data streams or finite data sets.
This diversity presents challenges with respect to 1) holistic, optimizable, intermediate representations, 2) efficient and scalable physical operators across all paradigms that can be mixed and matched, and 3) the combination of domain-specific and generic query languages that offers a sufficiently powerful yet optimizable interface to a data engineer.
Furthermore, the programming and reasoning about sensors and actuators in such a distributed, diverse setting entails a huge challenge with respect to both, scalability and ease of use.

To support diverse workloads in NES and create a large community with diverse users from different fields, we envision an \textit{easy-to-use} interface.
In particular, we attempt to allow users to choose their preferred programming environments and models without the need to take system-internals and performance implications into account.
To enable this diversity, we build on top of existing frameworks, such as Weld~\cite{palkar2017weld}, Arc~\cite{kroll2019arc}, Emma~\cite{Alexandrov2016}, and LARA~\cite{kunft2019optend} to represent diverse queries in a unified intermediate representation, our so-called \textit{Nebular-IR}. 
The Nebular-IR allows us to perform optimizations across operators, processing models, and language boundaries. 
The optimizations range from high-level optimizations on the operator plan level (e.g., placement, ordering, fusion \cite{hirzel2014acatalog}) to low-level optimizations on the instruction level (e.g., branch conversion across operators). 
One particular challenge for the Nebular-IR is to handle and optimize UDFs.
In particular, most data processing systems treat UDFs as black boxes and thus provide only basic optimizations to plans containing them.
However, in NES we first analyze UDFs to perform high-level optimizations on the IR (e.g., operator reordering~\cite{hueske2012opening}). 
After that, we fuse operators across UDF-boundaries and generate compact machine code.
This allows NES to achieve high code efficiency among different UDFs.

From a management point of view, centrally managing the system in a heterogeneous distributed setup introduces challenges from areas such as data collection, response time, and fault-tolerance.
To this end, NES provides a management view with a centralized, homogeneous interface, automatic distribution and parallelization, and means to adaptively detect and react to changes in the environment.
Although the management is performed centrally, parts of the system require a decentralized design.

By providing a central management view as well as an intermediate representation in NES, we support a diverse set of data processing workloads specified in different query languages and following different processing models.

\subsubsection{C5 - Constant Evolution under \\ Continuous Operation}
\label{sub:c5_constant_evolution_under_continuous_operation}
NebulaStream's goal is to support continuous operations while the topology and user workloads change constantly.
A particular challenge originates from a changing topology where new devices join the fog/WSN and existing devices get phased out or change their geo-spatial position.
Additionally, the workloads continuously change as users submit, update, or delete queries.
Furthermore, to enable time-sensitive processing, nodes must behave dynamically and autonomously during runtime, to capture and react to changes in velocity, volume, and variety.
Managing and reacting to changes in a robust way while the system is in continuous operation presents drastic challenges to the software architecture and fabric of an IoT data management system.

To support such a highly dynamic environment in NES, we apply the \textit{Autonomous Processing}, \textit{Dynamic Decisions}, and \textit{Incremental Optimizations} design principles.
First, NES equips the compute nodes with all necessary components to autonomously react to a wide range of situations.
We enrich the Node-EPs with several alternative routes and different options.
As a result, if a node detects changes in velocity, volume, or variety, it reacts dynamically at runtime. 
To this end, nodes require mechanisms to cope with a highly dynamic environment either locally, by interacting with nodes in the neighborhood, or by reaching out to a global coordinator.
The possible design space for these changes includes reduction of the sampling rate, dropping of packages, change in the operator order or algorithm, or rerouting of data streams.
Second, each software component in NES is designed to allow for the ever-changing network topology and query workloads and to handle some degree of bounded staleness.
We expect that this dynamicity will result in a complete redesign of many components and will require new algorithms and protocols.
In particular, we plan to incorporate the actor model \cite{DBLP:journals/pacmpl/BernsteinBBCFKK17} to capture the dynamic behavior between moving devices.
In this model, each device represents either a client, worker, source, or coordinator actor.
Using the actor model, we make sure that each device is always in a valid state and that each device can react to a wide range of events autonomously, e.g., lost connection or coordinator change.
We plan to use the actor model for coordination between actors, e.g., sending queries or reacting to node failures.
Due to the high message overhead of the actor model, we plan to offload data transfer to a more light-weight mechanism, e.g., ZMQ\footnotemark or RabbitMQ\footnotemark. 
\footnotetext[1]{https://zeromq.org/}
\footnotetext[2]{https://www.rabbitmq.com/}
Third, we apply incremental optimizations such that NES modifies a stateful execution plan of a running query in incremental steps rather than in one large change.
With each incoming or modified query as well as with each change in data velocity, volume, or variety, NES converges to the optimal NES-EP.
Furthermore, we introduce continuous feedback loops between the NES Coordinator and the NES Node Engines in different layers to enable a central management in a heterogeneous distributed setup.
In addition, NES re-optimizes the query execution based on dynamic changes in the workload and environment in an asynchronous process.
The trade-off between a centralized orchestration in a coordinator and decentralized decisions in the nodes remains an open research question for the future.
 
By defining feedback loops between its components and by performing changes incrementally and autonomously, we attempt to make NES resilient against constantly changing user workloads and network topologies.
Running AI/ML/Data Science algorithms based on data sets and streams produced by sensors in the IoT provides explanation models and prediction capabilities, which in conjunction with actuators, result in a feedback loop between the digital and the physical world.
Additionally, programming and reasoning about sensors and actuators in a distributed, diverse setting at the scale of the IoT provides a huge challenge both with respect to ease of use and scalability.

Overall, NebulaStream addresses all challenges of an IoT data management system presented in Section~\ref{subsec:nes_challenges} by combining existing approaches with new solutions. 
To this end, NebulaStream's goals are to handle heterogeneous and distributed data sources and formats, to utilize available resources efficiently, to cope with unstable network topologies, and to provide multiple query and processing models. 
We envision that NES's unique features make it an attractive platform for future IoT application scenarios.
\section{State-of-the-art Systems}
\label{sec:sota}
In this section, we group existing approaches and outline how they address  IoT data management challenges.

\vspace{1mm}
\subsection{Cloud-centric IoT data processing}
\label{subsec:cloud_rel_work}
The first group of approaches relies on the cloud to process IoT data centrally.
Mobile Cloud Computing (MCC) outsources data storage and processing from devices to the cloud.
In this scenario, a pool of sensors gathers and sends data directly to a cloud infrastructure for further processing ~\cite{aws_iot_analytics,azure_iot_hub}.
Example applications following this approach are camera surveillance~\cite{nest,netatmo}, wearable cognitive assistance~\cite{glass, hololens}, and smart city monitoring~\cite{fogConsortium2017visual, fogConsortium2017smartcity}.
As soon as data reach the cloud, common SPEs, such as Apache Flink~\cite{yang2017flink,piasecki2018Flink,sfikas2019Flink} and Apache Pulsar~\cite{kjerrumgaard2018pulsar2,bock2018pulsar1} process the incoming streams.
Based on this infrastructure, cloud providers offer services to deploy and manage data streams.

The cloud-centric processing of sensor data enables elastic scaling of compute and storage resources once data reach the cloud.
However, this neglects the resources provided by sensors and intermediate nodes (\textbf{C1},\textbf{C2}).
Although these systems offer fault-tolerance and dynamic scaling (addressing \textbf{C3},\textbf{C5}) in the cloud, they do not provide them across a unified sensor-fog-cloud environment.
In NES, we extend existing work in the area of stream processing to incorporate IoT specific requirements. In particular, we enable cross-paradigm optimization, in-network processing, and hardware-tailored code generation.

\vspace{1mm}
\subsection{Edge-Aware IoT data processing}
\label{subsec:cloud_rel_work}
With the concept of Mobile-Edge Computing (MEC), cloud providers address the limitations of cloud-centric approaches by implementing \textit{hub devices} to extend their IoT services \cite{Streaming_IOT_Survey,amazon_greengrass,azure_iot}.
Hub devices are placed at the edge of the fog topology and act as local control centers which are close to the sensors.
They gather data from attached sensors, perform simple processing steps, and do not require a stable connection to a cloud infrastructure.

Although MEC and MCC improve scalability with respect to the number of  sensors (addressing \textbf{C1}), they do not focus on efficient resource utilization across heterogeneous devices (\textbf{C2}).
In particular, hub devices do not enable cooperative processing across the whole topology.
Furthermore, these approaches offer fault-tolerance only between hub-devices and the cloud but still require a stable connection between sensors and the hub-device (partially addressing \textbf{C3}).
Additionally, these approaches do not address dynamic changes in the topology (\textbf{C3}).

Ryden et al.~\cite{ryden2014nebula} introduce a distributed data and resource management framework.
They leverage distributed in-situ data and computing resources on edge nodes only for batch processing.
Their system supports the combination of dedicated and voluntary resources under a unified infrastructure while ensuring high availability (addressing \textbf{C5}, partially addressing \textbf{C1}).
However, their framework neither exploits hardware heterogeneity for efficient code computation nor supports a multi-programming environment (\textbf{C2, C4}).
In NES, we support streaming queries in a unified sensor-fog-cloud environment that is able to exploit fog devices and sensors to optimize query execution in a holistic way.

\subsection{Fog-aware IoT data processing}
Two data processing systems utilize the fog as the underlying infrastructure.
O'Keeffe et al.~\cite{Frontier} propose Frontier, a distributed and resilient data processing system for fog devices.
Frontier aims to handle a large number of sensors and to achieve reliability.
To this end, it exploits the processing capability of the fog by distributing queries over a topology (addressing \textbf{C1}).
It replicates operators to neighboring nodes to recompute intermediate results and to cope with device failures (addressing \textbf{C3}).
However, Frontier does not address the efficient utilization of heterogeneous devices, diversity in programming environments, and adaptability to the constant evolution of the fog (\textbf{C2},\textbf{C4},\textbf{C5}).
Finally, it does not consider the exploitation of cloud resources.
 
Zhitao et al.~\cite{Streaming_IOT_Survey} extend Cisco's Connected
Streaming Analytics platform (CSA) for IoT processing.
CSA utilizes Cisco network hardware to enable in-network processing (partially addressing \textbf{C1},\textbf{C2}). 
However, CSA does not address potentially unreliable connections, the dynamic evolution of the fog, and provides only an SQL-like interface (\textbf{C3},\textbf{C5},\textbf{C4}).

In NES, we build on top of these approaches and combine the possible compute and storage capacities of the fog and the cloud.
Besides Frontier and CSA, additional research has been conducted on individual challenges in fog computing, which we will leverage in NES.
Janssen et al.~\cite{janssen2018scheduling} propose operator placement techniques to partition queries across a fog topology (addressing \textbf{C1}).
Park et al.~\cite{park2018StreamBoxTz} exploit special capabilities of IoT hardware to improve efficiency and security~(addressing \textbf{C2}).
Kang et al.~\cite{kang2017neurosurgeon} and Grulich et al.~\cite{grulich2018collaborative} propose solutions to partition the inference of deep neural networks across fog topologies to improve scalability (addressing \textbf{C1}).

\subsection{Data Processing in Sensor Networks}
Sensor networks (SNs) target a particular sub-area of the IoT~\cite{tinyDB,Cougar}.
In particular, these systems focus on distributed processing in a wireless sensor network~\cite{WSN_SURVEY}.
A major goal is resilience to intermittent and changing network connectivities.
To this end, sensor nodes form a network to transfer sensor values through multiple hops to a root node and perform in-network data processing.
Approaches in this area tackle efficiency (addressing \textbf{C2}) by optimizing the computation for battery lifetimes and enable filtering and aggregation queries over sensor data \cite{tinyDB}.
Moreover, they provide support for a dynamic execution environment (addressing \textbf{C5}).
However, these approaches do not support more complex and general workloads, which combine multiple queries, languages, and algebras (\textbf{C4}).
In addition, they do not provide strong fault-tolerance and correctness guarantees (\textbf{C3}).

In NES, we leverage concepts from sensor networks and integrate them seamlessly across the Sensor, Fog, and Cloud Layers, resulting in a unified environment.

\section{Conclusion}
\label{sec:conclusion}
In this paper, we introduced NebulaStream, a general-purpose, end-to-end data management system for the IoT.
We showed that current systems are not yet ready for the upcoming challenges of the IoT era.
We highlighted the system design of the NebulaStream platform and its design principles.
The goal of our envisioned design is to handle the heterogeneity, unreliability, and elasticity of a unified sensor-fog-cloud environment.
Furthermore, we revealed upcoming research challenges and outlined possible solutions.
Finally, we presented first results that motivate the need of a new system design for upcoming IoT applications.
With our \mbox{NebulaStream Platform}, we aim to enable emerging IoT applications in different domains.

\section{Acknowledgments}
This work was funded by the EU projects E2Data (780245), DFG Priority Program “Scalable Data Management for Future Hardware” (MA4662-5), FogGuru (Horizon 2020 under Marie Skłodowska-Curie grant agreement No 765452), the German Ministry for Education and Research as BBDC~II (01IS18025A), and by the German Federal Ministry for Economic Affairs and Energy as Project ExDra (01MD19002B).
Bonaventura Del Monte is partially funded by the German Ministry for Education and Research as Software Campus 2.0 (01IS17052).
We thank Julius Hülsmann for his support with the experiments on decentralized joins and Vianney de Cibeins for his support with the experiments on data reduction techniques.
Furthermore, we thank Eleni Tzirita Zacharatou and Xenofon Chatziliadis for the valuable input and discussions.

\bibliographystyle{abbrv}
\balance
\bibliography{sigproc}  
\balancecolumns

\end{document}